\documentclass[conference]{IEEEtran}

\IEEEoverridecommandlockouts
\usepackage{cite}
\usepackage{amsmath,amssymb,amsfonts}
\usepackage{algorithm}
\usepackage{algpseudocode}  
\usepackage{graphicx}
\usepackage{textcomp}
\usepackage{xcolor}
\usepackage[caption=false,font=footnotesize]{subfig}  
\usepackage{array} 
\def\BibTeX{{\rm B\kern-.05em{\sc i\kern-.025em b}\kern-.08em
    T\kern-.1667em\lower.7ex\hbox{E}\kern-.125emX}}
\begin{document}

\title{Dora: A Controller Provisioning Strategy in Hierarchical Domain-based Satellite Networks\\

\thanks{This work was supported by FDUROP (Fudan Undergraduate Research Opportunities Program) (24220), and National Undergraduate Training Program on Innovation and Entrepteneurship grant (S202510246451).}
}

\author{
Qiyuan Peng, Qi Zhang, Yue Gao, Kun Qiu \\
Institute of Space Internet, Fudan University, China \\
\{qypeng23, qizhang23\}@m.fudan.edu.cn, \{gao.yue, qkun\}@fudan.edu.cn
}

\maketitle

\begin{abstract}
The rapid proliferation of satellite constellations in Space-Air-Ground Integrated Networks (SAGIN) presents significant challenges for network management. Conventional flat network architectures struggle with synchronization and data transmission across massive distributed nodes. In response, hierarchical domain-based satellite network architectures have emerged as a scalable solution, highlighting the critical importance of controller provisioning strategies. However, existing network management architectures and traditional search-based algorithms fail to generate efficient controller provisioning solutions due to limited computational resources in satellites and strict time constraints. To address these challenges, we propose a three-layer domain-based architecture that enhances both scalability and adaptability. Furthermore, we introduce Dora, a reinforcement learning-based controller provisioning strategy designed to optimize network performance while minimizing computational overhead. Our comprehensive experimental evaluation demonstrates that Dora significantly outperforms state-of-the-art benchmarks, achieving 10\% improvement in controller provisioning quality while requiring only 1/30 to 1/90 of the computation time compared to traditional algorithms. These results underscore the potential of reinforcement learning approaches for efficient satellite network management in next-generation SAGIN deployments.

\end{abstract}


\section{Introduction}
In the coming decades, the Space-Air-Ground Integrated Network (SAGIN) is expected to experience exponential growth. Major satellite constellations such as Starlink (planning to deploy approximately 12,000 satellites with a potential expansion to 34,400) and Guowang (with plans submitted to the International Telecommunication Union (ITU) for a 13,000-satellite network) will collectively deploy tens of thousands of satellite nodes and aerial platforms \cite{b1}\cite{b2}. This rapid expansion will transform SAGIN into a massive and highly complex system, with the total number of network nodes potentially reaching hundreds of millions.

Such a dramatic increase in scale imposes stringent requirements and challenges on existing network management architectures. Traditional centralized control models based on ground stations (GS) \cite{b3}\cite{b4} struggle to handle the synchronization and data transmission needs of massive node populations. These architectures begin to fail at around 10,000 nodes, incurring exponential signaling overhead and high synchronization latency for Low Earth Orbit (LEO) satellites \cite{b27}. To overcome these limitations, hierarchical domain-based satellite network architectures \cite{b5}\cite{b6} have emerged as a promising solution, offering better scalability and adaptability. In these architectures, network is organized into multiple layers and domains, each with specific management responsibilities and autonomous capabilities. Given the large scale of SAGIN, the strategies for controller provisioning (including number, location and controlled members) significantly influence network management efficiency and fundamentally affect networks performance\cite{b7}.

Compared with flat network model, hierarchical domain-based satellite networks introduce greater dynamism. Due to the time-varying topology and dynamic traffic load of satellite networks, controllers must be configured in an online manner, as static or offline strategies cannot adapt to real-time changes in topology and traffic, which brings a restrict limitation of computation time. However, this real-time requirement conflicts with the limited computational capabilities of satellites. As a result, most hierarchical satellite networks \cite{b7}\cite{b8}\cite{b9} rely on ground stations to compute provisioning strategies and then transmit them to satellites. This offloading introduces additional latency—from 20 milliseconds (for satellites directly overhead) to 40 milliseconds (at the edge of the horizon)\cite{b30}—which negatively affects network responsiveness.

To provide a high-quality controller provisioning strategy with short computation time and lower computation ability requirement, first, we create a three-layer domain-based network management architecture and build a network evaluation model, which reformulates the controller provisioning problem as a optimization task with a well-defined objective function. Then we propose Dora, a controller provisioning model based on reinforcement learning, which has been successfully applied in satellite resource management\cite{b29}. Owing to the decoupled training and inference mechanism \cite{b10}, the model can be trained on ground stations equipped with high-performance computing resources and subsequently deployed on controller satellites for efficient real-time inference. Finally, we validated the algorithm’s performance. Experimental results demonstrate that Dora significantly outperforms benchmarks, spending only 1/30 to 1/90 computation time compared to traditional algorithm getting 10\% better performance of controller provisioning strategy. Otherwise, Dora exhibits strong scalability, maintaining high quality and low computation time in large scale network.

The contributions of this paper are outlined as follows:
\begin{itemize}
\item We propose a three-layer domain-based network management architecture and formerly define controller provisioning problem as a maximum optimization problem.
\item We propose Dora, a controller provisioning model based on reinforcement learning with short computation time and low computation ability requirement.
\item We compare Dora with other advanced controller provisioning algorithm validating the advantage of Dora.
\end{itemize}

The paper is organized as follows: 
Section~\ref{sec:background} briefly summarizes related work. 
In Section~\ref{sec:overview}, we overview the design of Dora. 
Section~\ref{sec:system_model} introduces the system model and defines the controller provisioning problem. 
In Section~\ref{sec:algorithm}, we propose the Dora algorithm to solve this problem. 
Section~\ref{sec:evaluation} presents performance evaluation, and 
Section~\ref{sec:conclusion} concludes the paper.

\section{Background} \label{sec:background}
\subsection{Network Management Architecture}
Typically, mainstream network management architecture can be divided into two categories: centralized controller architecture and hierarchical controller architecture\cite{b11}. 

In the first category, \cite{b3} \cite{b4} manage all satellite switches with a central controller deployed at the ground station. However, these flat network management architectures struggle with the vast scale of network, which brings high synchronizing overhead and long control delay reducing network performance. Also, subject to geography factors, ground station cannot be built all around the Earth. Therefore, satellite coverage is low leading to excessive control information overhead.

In the second category, \cite{b7}\cite{b12}\cite{b13} select suitable LEO satellites as controllers to manage surrounding LEO satellites. Since satellites in the same constellation have the similar motion, network achieves low and stable control latency. However, LEO controllers demand higher processing capabilities and inter-satellite link (ISL) performance, which is not currently feasible. \cite{b14} select MEO satellites and GEO satellites as controllers and super controllers, where former charges intra-domain items and latter charges inter-domain items. Transmission from MEO to GEO generates extra delay hindering real-time network management. Also, the complexity of hierarchical management reduces network reliability.

These limitations in existing architectures highlight the importance of efficient controller provisioning strategies, which we further discuss from the perspective of network optimization goals in the following section.

\subsection{Network Optimization Goals}
The existing network optimization goals can be roughly divided into three factors: network delay (the timeliness of data transmission and processing in the network), load balance (distributing network traffic evenly) and control overhead (resources consumed in implementing control strategies, including flow table update overhead, synchronization overhead, and predicted controller migration overhead). In \cite{b15}\cite{b16}, the optimization goal is to minimize the network delay between switches and controllers in the network. With the minimum network delay, users can get best use experience. \cite{b8}\cite{b17} set load balance as the optimization goal to make full use of network resources. \cite{b18} use minimum control overhead as the optimization goal. Through minimum control overhead, network can get global performance optimization.

Otherwise, there are lots of researchers who simultaneously consider multiple factors. \cite{b7} both considers load balance and control overhead as optimization goal, using a parameter $\lambda$ to adjust the weight. \cite{b19} combine network delay and link reliability to establish a multiobjective optimization problem. These diverse optimization objectives directly influence the design and performance of controller provisioning algorithms, which we review next.

\subsection{Controller Provisioning Algorithm}
Since the time-varying network topology and dynamic network load on each satellite, controller provisioning strategy need to be computed rapidly to ensure efficient network operation and adaptability. To address this problem, various studies have proposed different algorithm with low complexity. \cite{b7} propose an Approximate Regularization-based Online Algorithm (AROA) with $O(n^2)$ complexity and a Heuristic Regularization-based Online Algorithm (HROA) with $O(n^2lgn)$ complexity for large satellite constellations, which are both consumed to be global optimized. \cite{b18} proposes a CRG-based controller placement and assignment (CCPA) algorithm with a bounded approximation ratio, and design a lookahead-based improvement algorithm to further utilize the benefits of the predicted topology and traffic load known in advance.\cite{b8} proposes an two-step algorithm based on k-means and GA. 

However, all existing algorithms are search-based approaches, which typically demand significant computational resources. In contrast, model-based algorithms, such as reinforcement learning, can be pre-trained in advance. As a result, their deployment requires less computational power and significantly reduces runtime. Motivated by these advantages, we adopt a model-based approach and present our reinforcement learning-based strategy, Dora, in the following sections.

\section{Overview Design of DORA} \label{sec:overview}

\begin{figure*}[t]
\centering
\includegraphics[width=1.6\columnwidth, height=7.5cm]{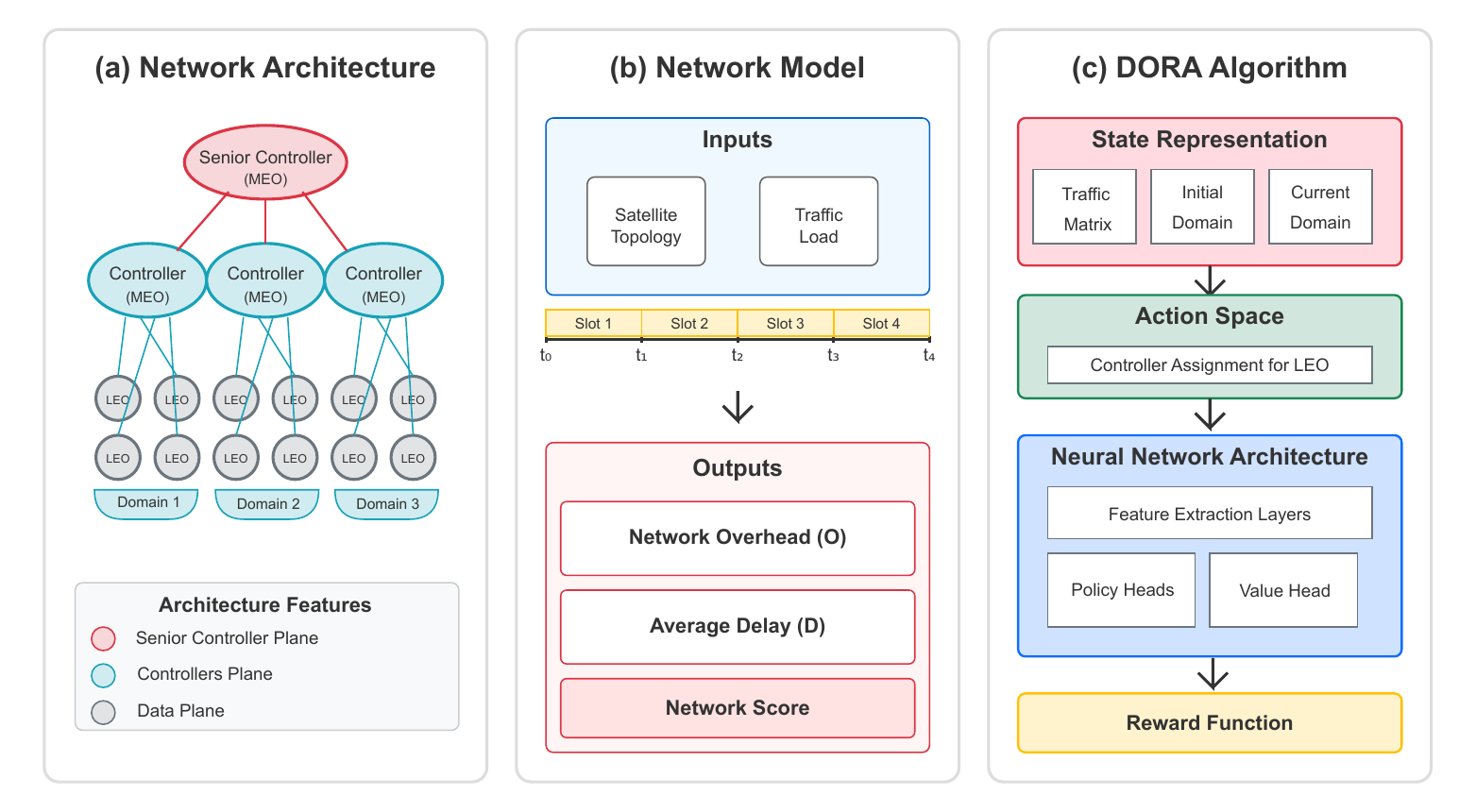}
\caption{Overview Design of Dora}
\label{fig:system_overview}
\end{figure*}

Fig. \ref{fig:system_overview} illustrates our proposed three-layer network architecture and the Dora framework. Our design encompasses three major components:

First, we propose a three-layer hierarchical domain-based network management architecture that divides the network into data plane (LEO satellites), controller plane (MEO controllers), and senior controller plane. This architecture provides both scalability to handle massive satellite constellations and adaptability to dynamic network conditions, while maintaining efficient control overhead.

Second, we formulate the controller provisioning problem as a multi-objective optimization problem that balances network overhead and response delay. We develop a comprehensive network model that captures time-varying satellite topology and traffic patterns, transforming the controller provisioning challenge into a well-defined mathematical optimization problem.

Third, we introduce Dora, a reinforcement learning-based controller provisioning strategy. By formulating the optimization as a Markov Decision Process and leveraging Proximal Policy Optimization, Dora generates high-quality controller provisioning solutions with significantly reduced computational requirements compared to traditional search-based algorithms.

\section{System Model and Problem Formulation} \label{sec:system_model}

\subsection{Three-Layer Hierarchical Network Architecture}
To manage the complexity of large-scale SAGINs, we adopt a three-layer hierarchical domain-based network management architecture, which is shown in Fig.\ref{fig:system_overview}(a). The \textbf{data plane} consists of LEO satellites for data transmission, grouped into domains, each managed by a controller. The \textbf{controller plane} comprises selected MEO satellites acting as domain controllers, responsible for intra-domain management and routing. The \textbf{senior controller plane} contains a high-performance MEO node that oversees inter-domain coordination and generates provisioning strategies in real-time. 

MEO satellites are selected for control tasks owing to their balance between computational capability and communication latency. They provide stronger processing power than LEO satellites, while ensuring lower latency to LEO nodes compared to GEO satellites or ground stations.

\subsection{Network Model}
We discretize time into slots $t = 0, 1, 2, \dots$ and assume network topology $G_t(S, E)$ remains stable within each slot. $S^L$ and $S^M$ denote sets of LEO and MEO satellites, with $E$ representing inter-satellite links. Domain assignment is modeled by $H_i$, the set of LEO nodes controlled by MEO controller $S^M_i$. 

Traffic is represented by matrix $T_t(i,j)$ and $f_t(i,j)$, indicating data volume and number of flow between $S^L_i$ and $S^L_j$ at time $t$. Transmission delay between satellite $i$ and $j$ is approximated by:

\begin{equation}
speed_{ij} = \frac{d_{ij}}{v_{light}},
\end{equation}
where $d_{ij}$ is link distance and $v_{light}$ is the speed of light. Path computation delays for intra- and inter-domain routing are denoted by $D^p_i$ and $D^p_*$, respectively.

\subsection{Network Overhead}
The total network overhead $O_t$ includes three components: synchronization overhead, flow table update overhead, and path computation overhead, which can evaluate the \textbf{overall system cost} from the company's perspective. These are given by:
\begin{align}
O^{\text{syn}}_t &= \sum_i Ts(i)_t \cdot \max_{j \in H_i} speed_{ij} \notag \\
&\quad + Ts(S^{M*})_t \cdot \max_i speed_{i*} \\
O^{\text{flow}}_t &= \sum_i \bigg( 
\sum_{\substack{j,k \in H_i}} T_t(j,k) \cdot speed_{ji} \notag \\
&\quad + \sum_{\substack{j \in H_i \\ k \notin H_i}} T_t(j,k) \cdot speed_{i*} \bigg) \\
O^{\text{path}}_t &= \sum_i D^p_i + D^p_*
\end{align}

Then, the total overhead is:
\begin{equation}
O_t = O_t^{syn} + O_t^{flow} + O_t^{path}.
\end{equation}

\subsection{Average Response Delay}
We define average response delay $D^{avg}_t$ as the weighted average of intra- and inter-domain delays, which can evaluate the \textbf{speed and quality of network} from the user's perspective. The intra-domain delay for controller $i$ is:
\begin{equation}
D^{intra}_{i,t} = Ts(S^{M*})_t \cdot speed_{i*} + D^p_i + \sum_{\substack{j,k \in H_i}} T_t(j,k) \cdot speed_{ji},
\end{equation}
and the inter-domain delay:
\begin{equation}
D^{inter}_{t} = \max_i D^{intra}_{i,t} + \sum_{\substack{j \in H_i, k \notin H_i}} T_t(j,k) \cdot speed_{i*} + D^p_*.
\end{equation}
Combining both:
\begin{equation}
D^{avg}_t = \frac{1}{\sum_{i,j} f_t(i,j)} \sum_{i,j} \left( D^{intra}_{i,t} + D^{inter}_{t} \right).
\end{equation}

\subsection{Problem Formulation}
To jointly optimize operator cost and user experience, we define a composite network score. Since in large-scale systems, relative improvements in normalized cost ($o_t = O_t/O_{t_{initial}}$) and delay ($d_t = D^{avg}_t/D^{avg}_{t_{initial}}$) are often marginal and confined to a narrow range near 1, our score function applies a logarithmic penalty that is highly sensitive in this regime:
\begin{equation}
    Score_t = (1 - \alpha) \cdot \frac{\log\left(1 - o_t\right)}{2} + \alpha \cdot \frac{\log\left(1 - d_t\right)}{2}
\end{equation}
where $\alpha$ balances overhead and delay. If $o_t,d_t \ge 1$, a negative constant penalty is applied to the score. And the controller provisioning problem is then formulated as:
\begin{equation}
\textbf{maximize} \quad Score_t.
\end{equation}

\section{Controller Provisioning Algorithm} \label{sec:algorithm}
To address this maximum problem, we first formalize the optimization problem as a Markov Decision Process (MDP)\cite{b20} and propose an algorithm DORA based on Proximal Policy Optimization (PPO)\cite{b21} to solve this problem.

\subsection{Formalization of MDP}

We formulate the controller provisioning problem as a MDP, defined by the tuple $(S, A, P, R, \gamma)$:

\textbf{State Space ($S$):}  
Each state $s_t$ encodes the system at time $t$, including:  
a normalized traffic matrix, a one-hot encoded initial domain allocation and current domain allocation.

\textbf{Action Space ($A$):}  
Each action can simultaneously reassign up to $K$ \emph{different} LEO satellites to new MEO controllers:\[
\mathcal{A} =
\bigl\{
a=\{(i_\ell,j_\ell)\}_{\ell=1}^{m}
\;\big|\;
1\le m\le K,\;
\bigr\},
\]
where each $(i_\ell,j_\ell)$ maps ${\rm LEO}_{i_\ell}\!\rightarrow{\rm MEO}_{j_\ell}$.

\textbf{State Transition ($P$):}  
Given action $a$, the system deterministically updates the domain allocation. Thus, $P(s' \mid s, a) = 1$ for the resulting $s'$.

\textbf{Reward Function ($R$):}  
Potential difference with penalty:\[
R(s, a, s') = \gamma Score_{t,s'} - Score_{t,s} - \lambda C_{a_t}.
\]
where $\lambda>0$ a step penalty weight and $C_{a_t}$ a step penalty.

\textbf{Discount Factor ($\gamma$):}  
A scalar $\gamma \in [0,1]$ that controls the trade-off between immediate and future rewards.

\begin{algorithm}[t]
\caption{Controller Provisioning Algorithm (Dora)}
\label{alg:dora}
\begin{algorithmic}[1]
\Require Satellite network, Scenarios $T$, Initial allocations $D$, Hyperparams: $\gamma$, $F_{switch}$
\State Initialize policy $\pi_\theta$ and value $V_\phi$ networks
\State Initialize experience buffer $B \leftarrow \emptyset$
\For{episode $= 1$ to $max\_episodes$}
    \If{episode $\bmod F_{switch} == 1$}
        \State Sample new scenario $(t, d_0) \sim (T, D)$ 
    \EndIf
    \State Get initial state $s_0$
    \For{step $= 1$ to $max\_steps$}
        \State $a_t \sim \pi_\theta(a|s_t)$ 
        \State Apply $a_t$, get reward $r_t$ and next state $s_{t+1}$
        \State Store transition $(s_t, a_t, r_t, s_{t+1})$ in buffer $B$
        \State $s_t \leftarrow s_{t+1}$
        
        \If{buffer $B$ is full}
            \State UpdateNetworks($\pi_\theta, V_\phi, B, \gamma$) 
            \Statex \Comment{Perform PPO update}
            \State Empty buffer $B \leftarrow \emptyset$
        \EndIf
    \EndFor
\EndFor
\State \Return Trained policy $\pi_\theta$
\end{algorithmic}
\end{algorithm}

\subsection{Algorithm Design}
Based on the MDP formulation, we propose Dora, a reinforcement learning algorithm using PPO to optimize satellite domain allocation. The training procedure is detailed in Algorithm 1. In each episode, the agent interacts with the environment to collect a batch of experiences. Once the buffer is full, we update the policy and value networks using the PPO algorithm\cite{b21}, which involves multiple epochs of updates on the collected data.

\begin{itemize}
\item \textbf{State Encoder}: To handle the vast state space, we first engineer a compact, feature-centric representation by aggregating key attributes for each satellite—such as position, traffic load, and controller assignments—into a linear-scale state vector. This structured state is then processed by a Graph Neural Network (GNN) encoder, which explicitly models the network's topology to extract a low-dimensional, relation-aware feature vector.
\item \textbf{Actor-Critic Structure}: We employ an Actor-Critic architecture where both the policy and value networks are built upon Multi-Layer Perceptrons (MLPs). These MLPs take the encoded state features to produce action probabilities and state-value estimations, ensuring stable and efficient learning.
\item \textbf{Training Strategy}: The model is trained using PPO. The agent trains on a single scenario (including the satellite topology and traffic matrix) for a fixed number $F_{switch}$ of episodes before switching to a new one, which effectively balances deep exploration within one context with broad generalization across many.

\end{itemize}

\section{Performance Evaluation} \label{sec:evaluation}
\subsection{System Setup}
The system is set up on Ubuntu 22.04.4 LTS, and the experimental computer is equipped with an Intel(R) Xeon(R) Gold 6418H CPU, 30 GB of RAM, 512 GB of SSD storage and RTX 4090 GPU. Satellites topology ($G_t(S,E)$) is created by Poliastro library\cite{b22} and Simplified General Perturbations Model 4 (SGP4) library\cite{b23}. In the experiment, we use LEO satellites in 550km altitude and MEO satellites in 8000km altitude to simulate satellite constellation. Traffic flow ($T_t,f_t(i,j)$) is stimulated with data created by Network Simulator 3 (NS3) library\cite{b24}. Otherwise, we use Border Gateway Protocol (BGP)\cite{b25} and Open Shortest Path First (OSPF)\cite{b26} to stimulate path computation delay ($D_i^p)$.

\subsection{Analysis of Training Dynamics}

\begin{figure}[t]
\centering
\includegraphics[width=1\columnwidth]{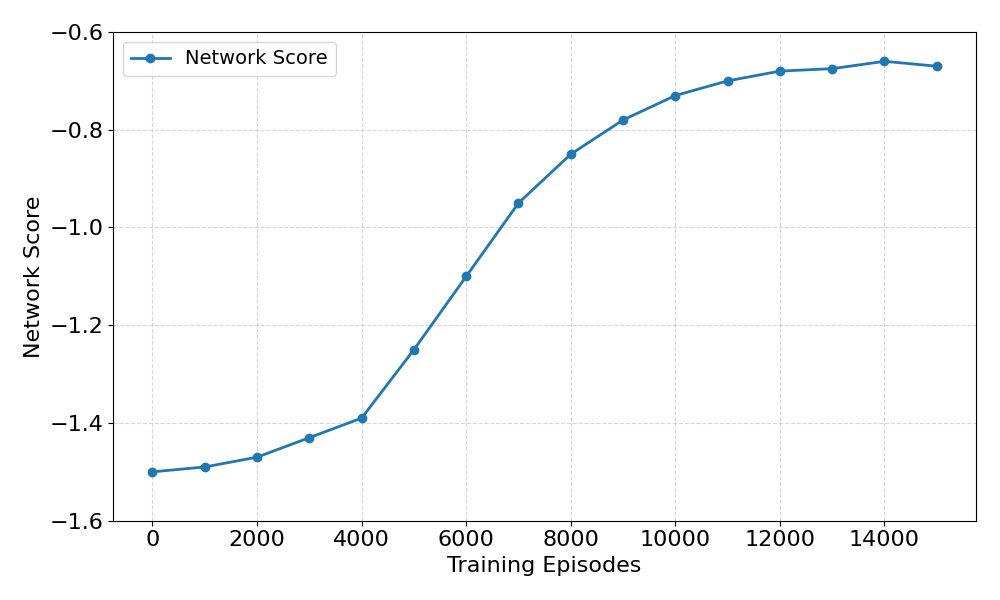}
\caption{Training Dynamics of Dora}
\label{fig:training}
\end{figure}

To clarify the learning trajectory, we track the model’s \emph{network score} over 15000 episodes using a constellation of 500 LEO and 20 MEO satellites with $\alpha = 0.5$. As shown in Fig.\ref{fig:training}, the network score rises from $-1.50$ to $-0.67$, a 55.3 \% improvement in the objective. In the early stage, the curve rises only modestly while the agent conducts broad exploration; through the middle stage, learning accelerates and the curve displays its steepest ascent; during the late stage, further gains diminish and the trajectory levels off, indicating that the controller-handover policy has converged. This three-phase progression mirrors the characteristic training dynamics of PPO, underscoring the effectiveness of the proposed model.

\begin{table}[t]
\caption{Performance under Different $\alpha$ Values}
\begin{center}
\begin{tabular}{|c|c|c|c|c|c|}
\hline
\textbf{$\alpha$} & \textbf{0.1} & \textbf{0.3} & \textbf{0.5} & \textbf{0.7} & \textbf{0.9} \\
\hline
\textbf{\textit{Enhanced Overhead}} & -0.49 & -0.55 & -0.58 & -0.65 & -0.79 \\
\hline
\textbf{\textit{Enhanced Delay}} & -0.90 & -0.80 & -0.76 & -0.72 & -0.69 \\
\hline
\textbf{\textit{Network Score}} & -0.53 & -0.63 & -0.67 & -0.70 & -0.70 \\
\hline
\end{tabular}
\label{tab:performance_alpha}
\end{center}
\end{table}

\subsection{Performance Analysis Under Different $\alpha$}
In this section, we evaluate the impact of the tunable parameter $\alpha$, which balances the weights of enhanced overhead ($\frac{\log\left(1 - o_t\right)}{2}$) and enhanced average delay ($\frac{\log\left(1 - d_t\right)}{2}$) in the optimization objective. Models were trained with $\alpha$ values ranging from 0.1 to 0.9 and tested under identical scenarios for fair comparison.
We adopted the same satellite constellation as in Subsection B to maintain consistency. 

As shown in Table \ref{tab:performance_alpha}, the delay increases as $\alpha$ increases, indicating that the model increasingly prioritizes delay during optimization. This trend aligns with our expectations and demonstrates that $\alpha$ effectively controls the trade-off between delay and overhead.

\subsection{Scalability Analysis of Dora}

\begin{figure}[t]
\centering
\includegraphics[width=1\columnwidth]{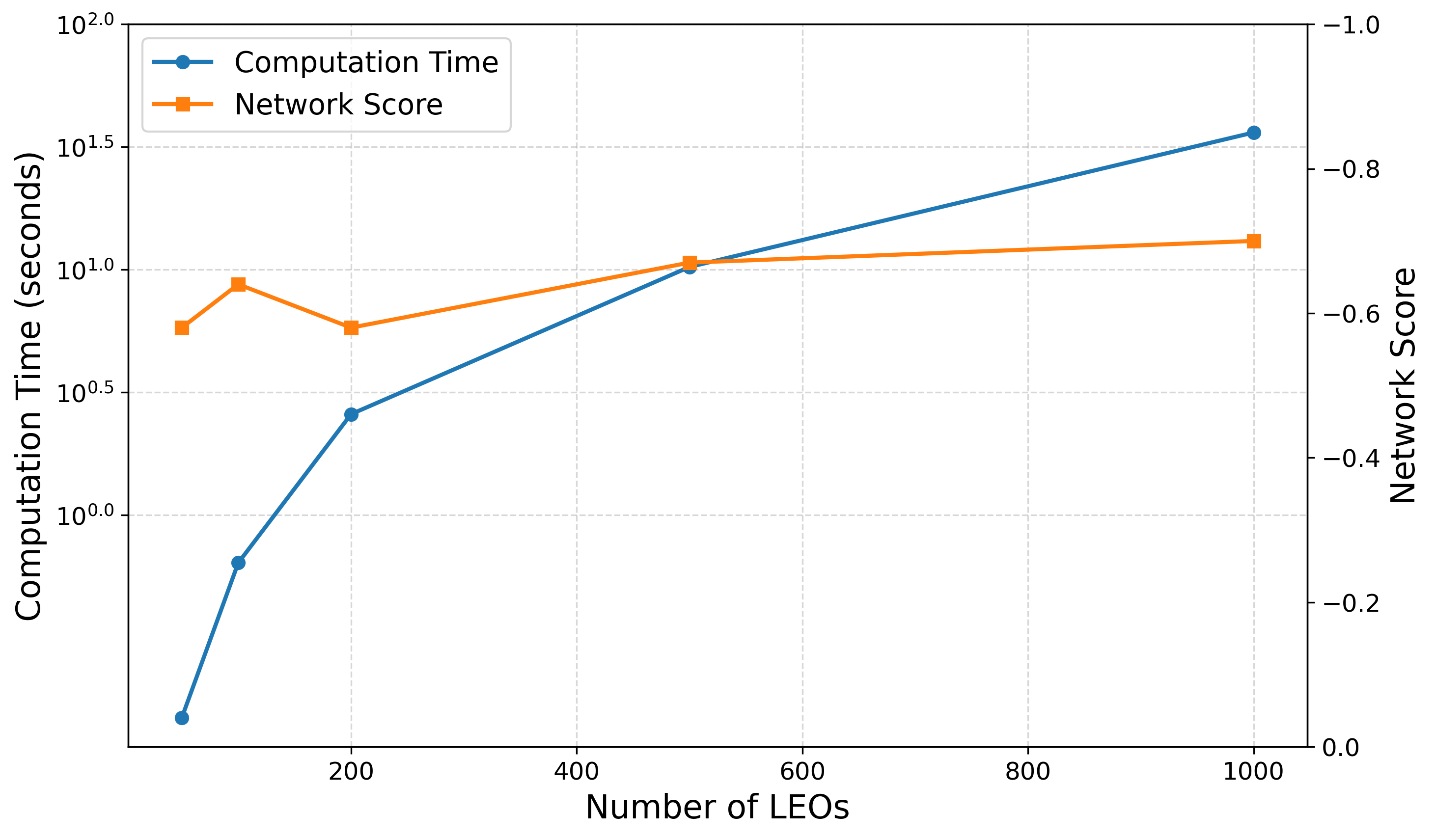}
\caption{Performance of Dora in different satellite constellation scale}
\label{fig:scale}
\end{figure}

In this section, we evaluate the scalability of Dora by training and testing models on satellite constellations ranging from 50 to 1000 LEO satellites. As shown in Fig. \ref{fig:scale}, Dora maintains consistently high network grades across all scales, demonstrating strong generalization to larger and more complex topologies without performance degradation. In terms of computational efficiency, Dora’s computation time increases polynomially with constellation size, approximately scaling with the first to second power of the satellite count. Even at 1000 satellites, the runtime remains practical for real-time deployment, confirming Dora’s feasibility for large-scale satellite systems.

\subsection{Performance Comparison with Other Algorithms}
In this section, we compare the performance of Dora with two related state-of-the-art algorithms: HROA \cite{b7} and an optimized genetic algorithm combined with K-means clustering (denoted as GA + K-means) \cite{b8}. All methods were evaluated under the same experimental setup, using the satellite constellation described in Subsection B.

\begin{figure}[t]
\centering
\subfloat[Network Quality\label{fig:dora_perf}]{%
  \includegraphics[width=0.45\linewidth]{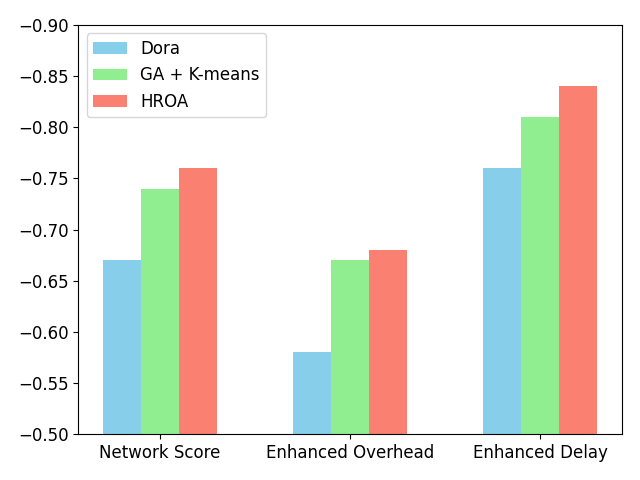}
}
\hfill
\subfloat[Computation Time\label{fig:traditional_comp}]{%
  \includegraphics[width=0.45\linewidth]{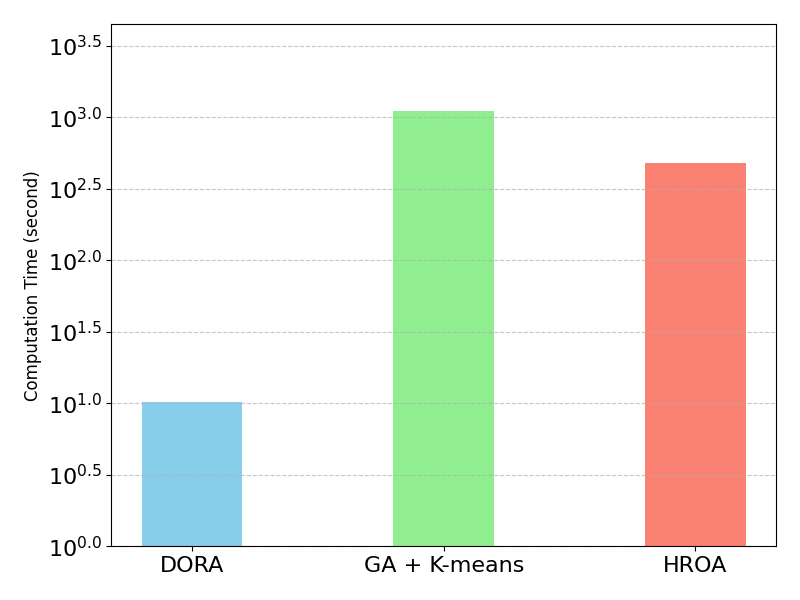}
}
\caption{Performance Comparison of Dora, GA + K-means and HROA}
\label{fig:algorithm_comparison}
\end{figure}

As demonstrated in Fig. \ref{fig:algorithm_comparison}, Dora achieves approximately a 10\% improvement in controller provisioning quality, considering both control overhead and transmission delay, compared to GA + K-means and HROA. In addition, Dora significantly outperforms both baselines in computational efficiency: its average computation time is 10.26 seconds, which is over 100 times faster than GA’s 1105.25 seconds and more than 45 times faster than HROA’s 482.27 seconds.

\section{Conclusion} \label{sec:conclusion}
In this paper, we addressed the critical challenge of controller provisioning in hierarchical domain-based satellite networks. We proposed a scalable three-layer network management architecture and formulated the provisioning task as a multi-objective optimization problem. To solve it, we introduced Dora, a reinforcement learning-based strategy designed to optimize network performance with minimal computation overhead. Extensive simulations show that Dora achieves 10\% higher provisioning quality while requiring only 1–3\% of the computation time compared to traditional algorithms. Even at large scales with 1,000 LEO satellites, Dora maintains near-optimal performance within 30 seconds of inference time.

Our results highlight the potential of reinforcement learning in addressing complex optimization problems in satellite networks. By decoupling the training and deployment phases, Dora enables efficient online decision-making with minimal computational overhead, making it a practical solution for future large-scale satellite constellations.

\end{document}